\documentclass[prl,12pt,citeautoscript,reprint,nofootinbib]{revtex4-1}
\usepackage{graphicx}
\usepackage{epsfig}
\usepackage{amsmath}
\usepackage{graphicx}
\usepackage{amssymb}
\usepackage{mathtools}
\usepackage{epsf}
\usepackage{epstopdf}
\usepackage{color}
\usepackage{subeqnarray}
\usepackage{mathrsfs}
\usepackage{ulem}
\usepackage[colorlinks=false, pdfstartview=FitV, linkcolor=green,
  citecolor=green, urlcolor=black, breaklinks]{hyperref}
\usepackage{natbib}
\usepackage{url}
\usepackage[utf8]{inputenc}
\usepackage[english]{babel}
\usepackage{float}
\usepackage{url}
\usepackage{breakurl}
\newcommand{\be}{\begin{equation}}
\newcommand{\ee}{\end{equation}}
\newcommand{\ben}{\begin{eqnarray}}
\newcommand{\een}{\end{eqnarray}}

\newcommand{\bsen}{\begin{subeqnarray}}
\newcommand{\esen}{\end{subeqnarray}}

\begin{document}

\title{Energy Exchange Calculations in a Simple Mechanical System to Investigate the Origin of Friction}

\author{María Luján Iglesias}\email{lujaniglesias@gmail.com}
\author{Sebastián Gonçalves}\email{sgonc@if.ufrgs.br}
\affiliation{Instituto de Física, Universidade Federal do Rio Grande
  do Sul, Caixa Postal 15051, 91501-970 Porto Alegre RS, Brasil}
\author{V.M. Kenkre}\email{kenkre@unm.edu}
\affiliation{Department of Physics and Astronomy,University of New Mexico, NM US}
\author{Mukesh Tiwari}\email{mukesh\_tiwari@daiict.ac.in}
\affiliation{Group in Computational Science and High Performance 
 Computing, DA-IICT, India}
\date{\today}

\begin{abstract}
The microscopic origin of friction is an important topic in science and technology. To date, noteworthy aspects of it remain unsolved. In an effort to shed some light on the possible mechanisms that could give rise to the macroscopic emergence of friction, a simple 1-d system of two particles is considered, one of them free but moving with an initial velocity, and the other confined by a harmonic potential. The two particles interact via a repulsive Gaussian potential. While it represents in a straightforward manner a tip—substrate system in the real world, no analytical solutions can be found for its motion. Because of the interaction, the free particle (tip) may overcome the bound particle (substrate) losing part of its kinetic energy. We solve the Newton’s equations of the two particles numerically and calculate the net exchange of energy in the asymptotic state in terms of the relevant parameters of the problem. The effective dissipation that emerges from this simple, classical model —with no ad hoc terms— shows, surprisingly, a range of rich, non-trivial, behavior. We give theoretical reasoning which provides a satisfactory qualitative description. The essential ingredient of that reasoning is that the transfer of energy from the incoming particle to the confined one can be regarded as the source of the emergent dissipation force—the friction experienced by the incoming particle.
\end{abstract}
\maketitle

\section{Introduction}
Energy dissipation, the statistical, macroscopic manifestation of fundamental interactions among particles at the atomic scale is known to give rise to friction. While considerable progress has been made in the fundamental understanding of its origins~\cite{Krim2002,BENNEWITZ200542,Buldum1997} during the last thirty years, the elucidation of the detailed microscopic mechanisms behind it is still an open problem—one which is an important one from a practical point of view~\cite{Persson2000,Krim2002}. 

The introduction of new experimental tools made the nanometer and the atomic scales accessible to tribologists, and gave rise to the field of nanotribology~\cite{Holscher2008}. The investigation of friction between surfaces at the atomic scale, such as an adsorbed mono-layer or an atomic force tip over a perfect crystalline substrate, is of high interest to nanotribology. The strategy is to understand first the two particles contact problem and then to use it to study macroscopic friction via the use of statistical mechanics. 

Dry sliding friction between atomically flat sliding surfaces, commensurate or incommensurate, is the most fundamental type of friction in tribology~\cite{Buldum1997}. While it is the simplest and most ubiquitous manifestation of electromagnetic forces between atoms and molecules at a macroscopic level, it is also the one less understood from first principles. In dry friction, many interesting and complex physical phenomena are involved, such as adhesion, atom exchange, elastic deformation and plastic deformation. The understanding of these phenomena could allow us to develop ways of controlling the mechanisms of friction and thereby to reduce the loss of energy in ordinary operations. 

In the past three decades, several theoretical models for atomic friction, based on the so-called Prandtl-Tomlinson~\cite{Prandtl1928,Prandtl1904,Prandtl1920,Tomlinson1929} and Frenkel-Kontorova~\cite{Kontorova:431596,Frenkel:431595,braun2004frenkel} models, have been proposed. The advantage of these models lies in the fact that they are simple yet retain enough complexity to exhibit interesting behavior~\cite{Buldum1997,Goncalves2004,Goncalves2005,Fusco2005,Tiwari2008,Neide2010,Apostoli2017}. Such models have displayed non-trivial features arising from the fact that the dissipation of energy and the resulting emergence of the force of friction are consequences of nonlinear mechanisms. However, in all those models, dissipation terms have been introduced in an explicitly external fashion. Even when the resultant friction is non-trivial, part of them is already set up from the beginning. Indeed, they all have in common the use of a Stokes or viscous dissipation term, proportional to velocity, as a way to simulate the effect of electronic friction which cannot be included from first principles like the vibrational part. This ad hoc inclusion of an a priori dissipation term is certainly a weakness of these models. In the present contribution, we refrain from employing such ad hoc artifacts, and  study the effective dissipation experienced by a particle moving freely but constrained by interaction with a harmonic oscillator which represents a substrate. The particle itself represents a sliding surface or a tip, moving past another one. 

Versions of the present model have been studied some time ago in connection with other problems different from the microscopic origin of friction. Secrest~\cite{Secrest1969}, for example, studied the energy exchange in atomic collisions between an incoming particle and a diatomic molecule represented by a harmonic oscillator. Although the system is closely related to the one presented here, the focus was on the energy transmitted to the dimer and the incoming particle was not allowed to continue because of the potential. Storm and Thiele~\cite{Storm1973} use a similar model but in a quantum framework to obtain the transition probabilities related to molecular collision in gases. Teubner~\cite{Teubner2005} studied the linear collision of a classical harmonic oscillator with a mass point in the high-frequency region as a way to approach the more complex quantum mechanical problem of chemical reactions. Also, Mavri et al.~\cite{Mavri1994} analyzed the inelastic collisions of a classical particle with a five level quantum harmonic oscillator, using density matrix evolution in a pseudo classical approach to the chemical reaction problem. All these four contributions have in common the one dimensional simplification and their focus is in chemical reaction problems.

Our goal here is to bring some understanding of the microscopic mechanisms of energy dissipation between two particles: one representing the substrate and the other representing the adsorbate or the tip of the microscope. At the most fundamental and specific level, we are able to observe the systematic loss of kinetic energy from the adsorbate to the vibration of the substrate, without resorting to an ad hoc dissipation term.

\section{Model}
\begin{figure}[!h]
\centering \includegraphics[width = 1.\columnwidth]{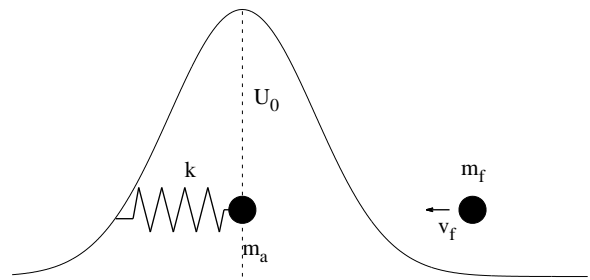}
\caption{Schematic representation of the two particle model.}
\label{fig:model}
\end{figure}

The model consists of two particles: a mass that is thrown to move past another one, which is attached to a spring (Fig.~\ref{fig:model}).  The interaction between the particles is nonlinear and short-ranged. For simplicity, the motion is considered in one dimension, but the free particle can overcome the bound one because the interaction potential between them has a finite maximum. This seemingly unreal situation is a simplified model of a tip moving on top of substrate, where the transverse movement is not taken into account. The tip is represented by a single free particle and the substrate by a harmonic oscillator. As shown in Fig.~\ref{fig:model} the free particle is impelled from a region where it is not affected by the presence of the oscillator, until eventually it interacts with it. This local interaction is repulsive, short ranged, and with a finite maximum; for simplicity a Gaussian is used for the interaction potential.

Therefore the proposed model is represented by the Hamiltonian 
\be 
H =\frac{p_{f}^{2}}{2m_{f}} + \frac{p_{a}^{2}}{2m_{a}} + \frac{1}{2}m_{a}\omega^{2}x_{a}^{2} + U(x_{f},x_{a})
\label{eq:model}
\ee
where $p_{f},~x_{f},~p_{a},~x_{a}$ are the momentum and position of the free particle and the oscillator respectively. The masses of the two are unequal and represented by $m_{f}$ and $m_{a}$ respectively. The oscillator frequency is $\omega = \sqrt{k/m_a}$, where $k$ is the spring constant. As mentioned earlier, the interaction potential $U(|x_{f}-x_{a}|)$ is assumed to be a Gaussian with a maximum value $U_{0}$, the height of the potential, and a width $\sigma$, which characterizes the typical range or length of the interaction. From the Hamiltonian in Eq.~(\ref{eq:model}), the following equations of motion are derived:
\ben
\label{Eq:newton_p}
\ddot{x}_{f} &=&
\left(\frac{2U_{0}}{m_{f}\sigma^{2}}\right)(x_{f}-x_{a})e^{-\frac{(x_{f}-x_{a})^{2}}{\sigma^{2}}}\\
\label{Eq:newton_osc}
\ddot{x}_{a} &=& -\omega^{2}x_{a}-
\left(\frac{2U_{0}}{m_{a}\sigma^{2}}\right)(x_{f}-x_{a})e^{-\frac{(x_{f}-x_{a})^{2}}{\sigma^{2}}}\;,
\een
which have five parameters. By defining dimensionless variables $q =
x/\sigma$ and $\tau =\sqrt{\frac{U_{0}}{m_{f}\sigma^{2}}}~t$, and
writing the equations of motion in terms of them,
\ben \ddot{q}_{f} &=&
2(q_{f}-q_{a})e^{-(q_{f}-q_{a})^{2}}\\ \ddot{q}_{a}&=&
-\frac{\Gamma^{2}}{\mu}q_{a} -
2\frac{(q_{f}-q_{a})}{\mu}e^{-(q_{f}-q_{a})^{2}} \;,
\een 
the number of free parameters can be reduced to two: the ratio of the masses of the two objects $(\mu = \frac{m_{a}}{m_{f}})$, and a dimensionless frequency $(\Gamma = \sqrt{\frac{k\sigma^2}{U_0}})$. In the following, we focus on these two parameters, the relative mass and the frequency, work with Eqs.~(\ref{Eq:newton_p}) and (\ref{Eq:newton_osc}), and characterize the behavior of the system in terms of those two parameters. We have chosen the interaction potential to be Gaussian for two reasons. First, it is short-ranged and thereby captures the essential features of the interaction between the “substrate” and “adsorbate”. Second, it makes the problem nonlinear with no closed analytic solution, thereby introducing sufficient richness in the dynamics

\section{Results}
\subsection{Free particle-Non Oscillator Interaction}
To underline the richness in behavior of the complete system, especially on the role of the frequency of the oscillator, we first set $\omega=0$ ($k=0$) in Eq.~(\ref{Eq:newton_osc}). By defining $x = x_{f} - x_{a}$ and $X = \frac{m_{f} x_{f}+ m_{a}x_{a}}{m_{f}+m_{a}}$ as the internal and center of mass coordinate respectively in Eqs.~(\ref{Eq:newton_p}) and (\ref{Eq:newton_osc}), we obtain the following equations:
\ben
\label{eq:intco}
\ddot{x}&=&
\left(\frac{2 U_{0}}{\sigma^{2}}\right)\left(\frac{1}{m_{f}} +
\frac{1}{m_{a}} \right)x e^{-x^{2}/\sigma^{2}}\\
\label{eq:com}
\ddot{X}&=& 0 \;.
\een
By assuming that the two particles are initially well apart from each other, that the free particle starts with velocity $-v_0\; (v_0 >0)$ moving to the bound mass, which is initially in equilibrium at rest, the following conservation equations can be derived from Eqs.~(\ref{eq:intco}) and (\ref{eq:com})
\ben
\label{eq:intcoord}
\frac{\dot{x}^{2}}{2} + U_{0}\left(\frac{1}{m_{f}} +
\frac{1}{m_{a}}\right)e^{-x^2/\sigma^{2}} &=& \frac{v_0^2}{2}\\
\label{eq:centerofmass}
\dot{X}(t)=\dot{X}(0)=-\frac{v_0 m_{f}}{m_{f}+m_{a}} \;\cdot
\een
The asymptotic relations between the velocities, when one of the particles has gone far away from the interaction region, {\it i.e.} the potential term goes to zero are: 
\ben
\label{eq:vel_int_coord}
v_{f}(\infty)-v_{a}(\infty)=\pm v_{0}\\
\label{eq:vel_int_coord2}
v_{f}(\infty)+\frac{m_{a}}{m_{f}} v_{a}(\infty)= -v_{0} \;.
\een
Combining Eq.~(\ref{eq:vel_int_coord}-\ref{eq:vel_int_coord2}), the minus sign in Eq.~(\ref{eq:vel_int_coord}) gives $v_a(\infty)=0$ and $v_{f}(\infty) = -v_0 = v_{f}(0)$: the incoming particle passes through the fixed one, moving away with the same asymptotic velocity it had initially, while the bound one stays at rest after the interaction.  Such asymptotic situation, in which the incoming particle does not loose energy after the interaction, happens whenever the initial kinetic energy of the incoming particle is larger than the maximum of the interaction potential energy, {\it i.e}:
\ben
|v_f(0)| > v_{cr} = \sqrt{2 U_{0}\left(\frac{1}{m_{f}}+\frac{1}{m_{a}}\right)} \;.
\label{eq:vcr}
\een
The plus sign in Eq.~(\ref{eq:vel_int_coord}) gives $v_{a}(\infty) =\left(\frac{2m_{f}}{m_{f}+m_{a}}\right)v_{f}(0)$ and ${v}_{f}(\infty) = \left(\frac{m_{f}-m_{a}}{m_{f}+m_{a}}\right)v_{f}(0)$ as solutions. Therefore, $v_f(\infty)$ can be positive, zero, or negative depending on the relation between the two masses. 

\begin{figure}[!h]
\centering
\includegraphics[width = 1.\columnwidth]{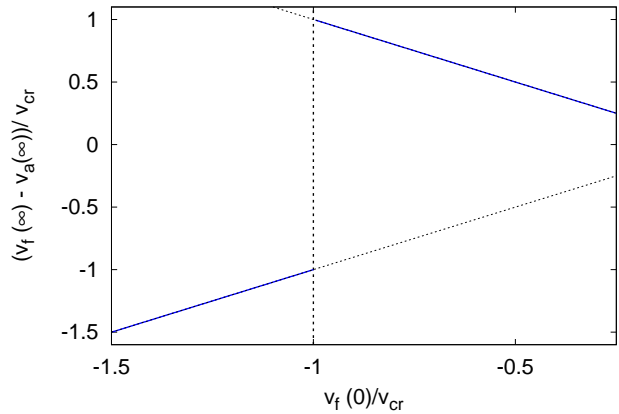}
\caption{Final velocity of the incoming particle as a function of its initial velocity normalized to the critical velocity $v_{cr}$ for $k=0$. The dashed lines show the analytical expression and the solid lines are obtained from numerical simulations. The other parameters are $U_0=1$ and $\sigma=1$. $m_a$ is initially at rest ($v_a(0)=0$) and far from the incoming particle $m_f$ ($v_f(0)=-v_0 < 0$).}
\label{Fig:vel_k_0}
\end{figure}

This is the well-known case of elastic collision which, in the center of mass reference system, has a unified description for the 
three cases, and as such has no direct relation with the friction problem we study. However, as we will observe in the next section, this understanding allows us to gain insight into the energy lost by the free particle. Figure~\ref{Fig:vel_k_0} shows a comparison between the numerical simulation and the analytical result for the final velocity of the free particle as a function of its initial velocity for the case $m_a/m_f = 2$.

\subsection{Free Particle-Oscillator Interaction}
\begin{figure}[!h]
\centering
\includegraphics[width = .7\columnwidth]{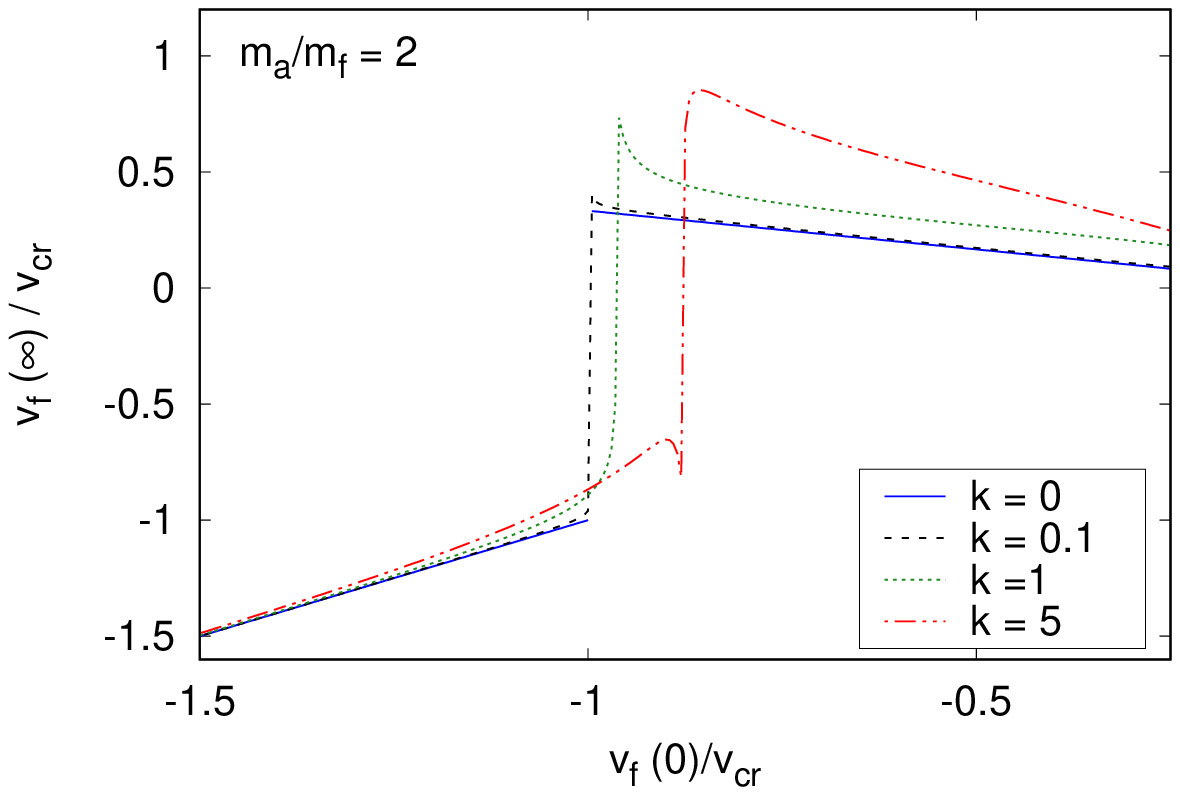} 
\includegraphics[width = .7\columnwidth]{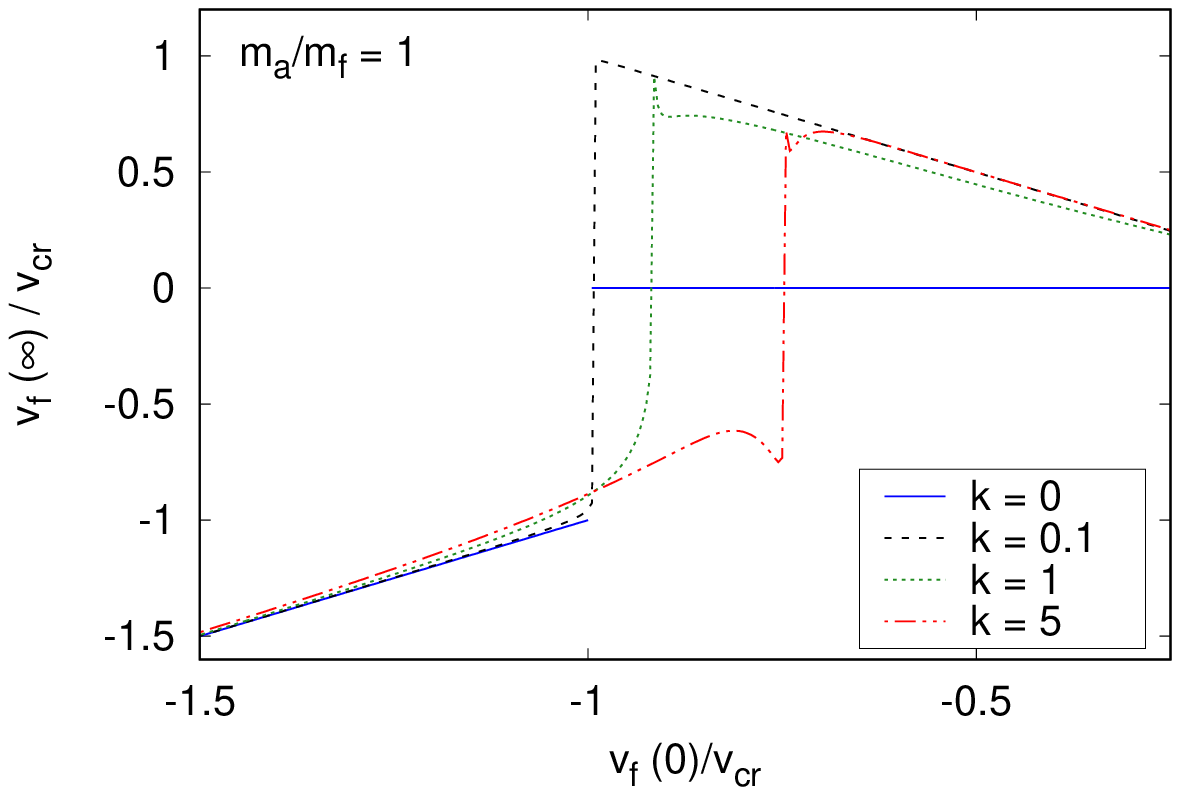} 
\includegraphics[width = .7\columnwidth]{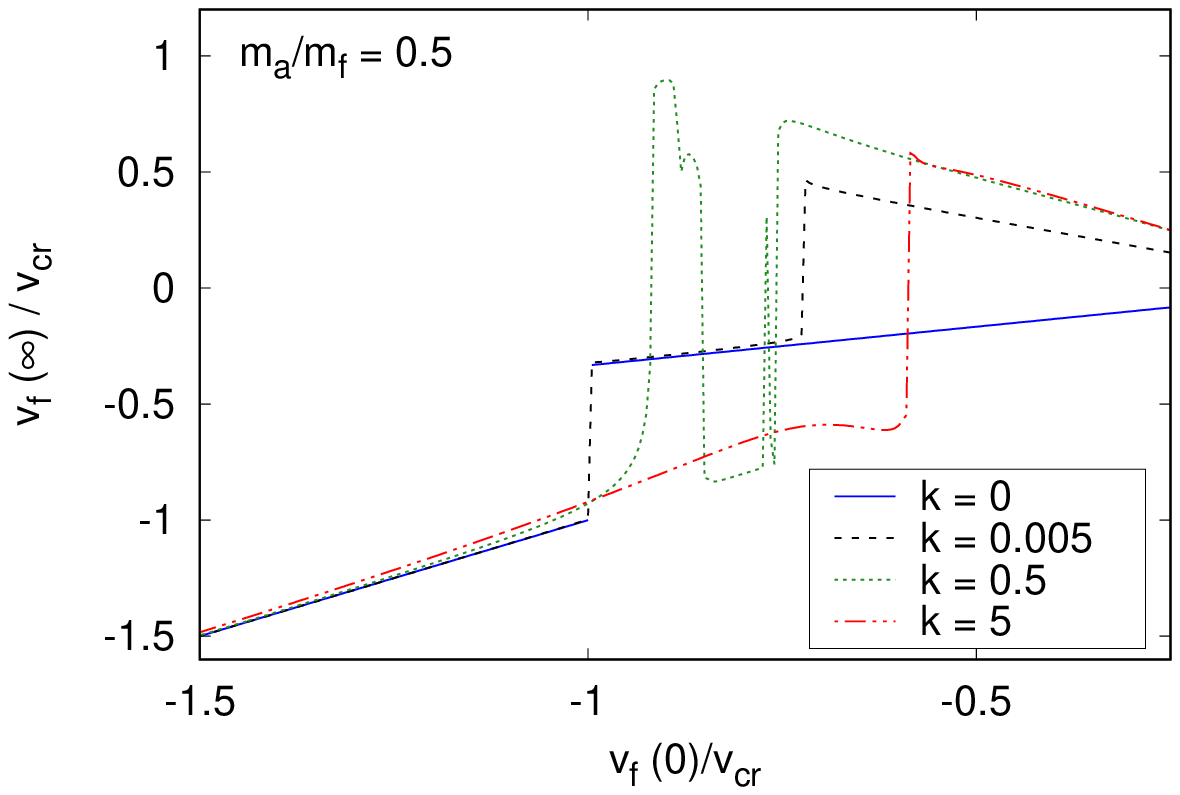}
\caption{Final velocity of the incoming particle, $v_{f}(\infty)/v_{cr}$, as a function of its initial velocity, $v_{f}(0)/v_{cr}$ (both in terms of the critical velocity) for the $(k\neq 0)$ case. The other parameters of the system are $U_0=1$ and $\sigma=1$, $m_a$ is initially at rest ($v_a(0)=0$) and far from the incoming $m_f$ particle ($v_f(0)=-v_0 < 0$).  Solid (blue) lines show correspond to $(k=0)$ cases and have been drawn for reference.}\label{Fig:slow_fast_osc}
\end{figure}

\begin{figure}[!h]
\centering
\includegraphics[width = 1.\columnwidth]{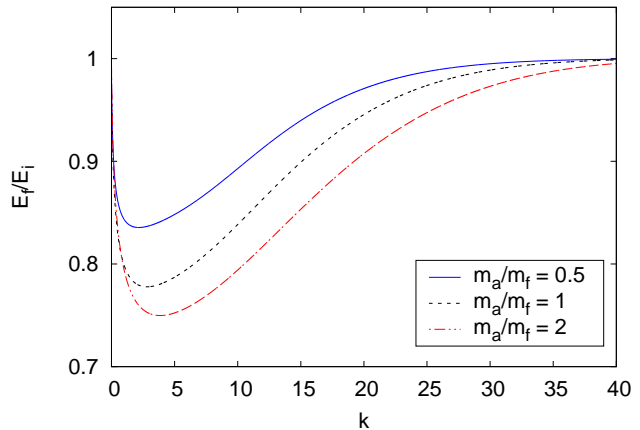}
\caption{Energy ratio vs $k$ for three different mass ratios. The
  initial velocity of the free particle is $|v_f(0)| = 2$.}
    \label{Fig:ene_vs_k}
\end{figure}

Binding the second particle to the equilibrium position of the harmonic potential $(k>0)$ leads to a complex relationship between the final and initial velocity of the free particle. In Fig.~\ref{Fig:slow_fast_osc}, for three different mass ratios $(\frac{m_{a}}{m_{f}} = 2,\;1$, and $0.5)$  and a range of $k$ values,	the complex relationship between the final and initial velocity is demonstrated. It is evident that the elasticity of the spring also influences the relationship significantly. For large initial velocities, independent of the mass ratio and strength of the spring, the free particle moves away from the interaction region with no loss of energy. For sufficiently small initial velocities, the free particle bounces back, losing some energy to the oscillator which depends on both its initial velocity and the spring constant. Both these situations are not of much interest because they have no implications on the friction
problem. Rich and non trivial behavior is observed in the intermediate velocity range, between the $k=0$ critical velocity and a new critical velocity that seems to depend on both $k$ and the mass ratio. Independent of the mass ratio, qualitative similarity can be observed between different panels in Fig.~\ref{Fig:slow_fast_osc}. The new critical velocity is observed to decrease as the values of $k$ increases. This is expected since in the limit of very large $k$, the problem reduces to that of a particle moving under the influence of a fixed potential. The ratio between the critical velocities for the two extreme limit $\frac{v_{cr}|_{k\rightarrow0}}{v_{cr|k\rightarrow\infty}} = \sqrt{1+\frac{m_f}{m_{a}}} $, indicates a shift to smaller $|v|$ values as $k$ increases. The extreme case of $k\rightarrow\infty$ is also not of any interest, since there is no emerging ``friction'' as the energy of the free particle is conserved. The exchange of energy between the free and the bound particle is therefore expected to show a a maximum at some intermediate value of $k$, and as observed in Fig.~\ref{Fig:ene_vs_k}, in addition to $k$ also depends on the mass ratio. 

Another general behavior observed in Fig.~\ref{Fig:slow_fast_osc} is that as the mass of the free particle ($m_f$) increases, relative to the bound mass, the change in the critical velocity becomes more sensitive to the stiffness of the bound particle, $k$. More precisely, the shift as a function of $k$ is more dramatic for larger value of the mass $m_f$. A peculiar situation with more than one jump, for intermediate $k$ values is observed when the free particle is more massive than the bound one (Fig.~\ref{Fig:slow_fast_osc}~lower panel), and is discussed in the following section.

\subsection{Soft \texorpdfstring{$k$}{Lg} (slow oscillations)}
 
\begin{figure}[!h]
\centering \includegraphics[width = 1.\columnwidth]{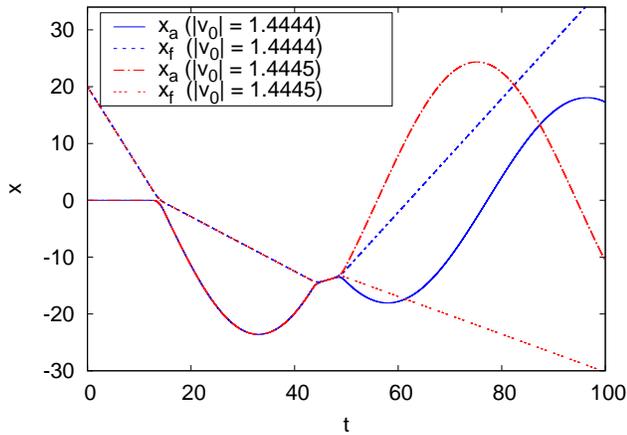}
\caption{Position of the free particle (dashed lines) and the oscillator (solid lines) for reflection (blue lines) and transmission (red lines) of the free particle after interaction with the oscillator. For the reflected case the initial velocity of the free particle is $|v_{f}(0)| = 1.4444$ and for the transmitted one, it is $|v_{f}(0)| = 1.4445$. The value of $k$ is taken to be 0.005 and $m_{a}/m_{f}=0.5$.  All other parameters are the same as in Fig.~\ref{Fig:slow_fast_osc}.}\label{Fig:trans_ref}
\end{figure}

For small values of $k$, the lower panel of Fig.~\ref{Fig:slow_fast_osc} ($m_a/m_f=0.5$) displays a small window in the transmission region of the free particle, in which it loses a significant amount of its energy to the oscillatory mode. Beyond this window, the free particle is reflected back with hardly any energy loss to the oscillator. Further, as can be seen in Fig.~\ref{Fig:trans_ref} the transition between transmission (red curves in Fig.~\ref{Fig:trans_ref}) and reflection (blue curves in Fig.~\ref{Fig:trans_ref}) is highly sensitive to initial conditions. That is, for initial velocities which differ in less than 1 in 10$^4$, the motion of free particle changes from reflection to transmission. For both the situations, however, we observe in Fig.~\ref{Fig:trans_ref} that the two particles interact with each other twice before the free particle leaves the interaction region. This behavior is observed over a range of parameter values. For small values of $k$, we first attempt to obtain an approximate estimate of the initial velocity for which the behavior changes from transmission to reflection. Since the two collisions are well separated in time, the analysis of the previous section for $k=0$ can be extended to the two collision observed here for $k \approx 0$.  The approximation considers that the two particles move independently ($m_f$ moves at constant velocity and $m_a$ is subject to the harmonic potential) except at the instant of collisions, for which the results of the previous section can be applied.

\begin{figure}[!h]
\centering
\includegraphics[width = 1.\columnwidth]{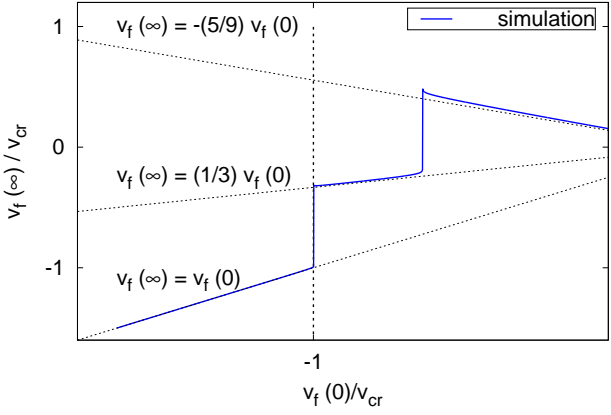}
\caption{Comparison of the results obtained from numerical simulations (solid line) with the approximate two collision analysis (dashed line) for slow oscillations. The value of $k=0.005$ and $m_{a}/m_{f}= 0.5$.}\label{Fig:slow_osc}
\end{figure}

Since the initial velocity of the free particle is less than the critical velocity $(v_{cr})$ defined in Eq.~(\ref{eq:vcr}), immediately after the first collision $(t>t_{1})$ the velocities of the two particles are:
\ben
\label{eq:velpcol1}
v_{f}(t_1^+) &=& \frac{m_{f}-m_{a}}{m_{f}+m_{a}}v_{f}(0)\\
\label{eq:velosccol1}
v_{a}(t_1^+) &=& \frac{2m_{f}}{m_{f}+m_{a}} v_{f}(0) \;,
\een
where, $t_i^{\mp}$ denote the time immediately before and after the collision.  Assuming that after this collision, the particles move independently, their positions are given by:
\ben x_{f}(t > t_1) &=& v_{f}(t_1^+) t\\
\label{eq:pospcol1}
x_{a}(t > t_1 )&=& \frac{\sin(\omega t)}{\omega} v_{a}(t_1^+) \;.
\label{eq:pososccol1}
\een
The instant of the next collision $t=t_{2}$ can be determined by setting $x_{f}(t_2)=x_{a}(t_2)$. From Eqs.~(\ref{eq:velpcol1})-(\ref{eq:pososccol1}) we obtain the following relation for $t_2$,
\ben
sinc (\omega t_{2}) = \frac{m_{f}-m_{a}}{2m_{f}} \;.
\label{eq:sinc} 
\een
Transfer of energy between the two is only possible if the velocity of the internal coordinate at the time of collision is less than the critical velocity $(v_{cr})$.  Using Eqs.~(\ref{eq:intcoord}), (\ref{eq:velpcol1}), and (\ref{eq:pososccol1}), the condition on initial velocity for which the behavior of the free particle changes from transmission to reflection is
\ben v_{f}(0) =
\frac{v_{cr}(m_{f}+m_{a})}{m_{f}(1-2\cos(\omega t_{2})) - m_{a}}\;.
\een
The instantaneous velocity after the second collision can be obtained from conservation of energy and momentum and is given by
\ben
\label{eq:velpcol2}
v_{f}(t_2^+)= \frac{2m_{a}}{m_{a}+m_{f}} v_{a}(t_2^-) +
\frac{m_{f}-m_{a}}{m_{a}+m_{f}} v_{f}(t_2^-)\\
\label{eq:velosccol2}
v_{a}(t_{2}^+)=\frac{m_{a}-m_{f}}{m_{a}+m_{f}}v_{a}(t_{2}^-) +
\frac{2m_{f}}{m_{a}+m_{f}}v_{f}(t_{2}^-) \;.
\een
Figure~\ref{Fig:slow_osc} shows a comparison between the approximate two collision theory and numerical results for $m_{a}/m_{f} = 0.5$ and $k=0.005$. For these parameter values, the different behaviors in the final velocity are distinctly observed and the two collision approximation is able to provide an excellent fit to the numerical results. There is another critical velocity (the lower bound in Fig.~\ref{Fig:slow_osc}, or the second discontinuity). When the free particle's velocity is larger than this critical velocity, the two particles move with almost the same velocity for a while, as if they are stuck together until the elastic force is enough to pull the bound one back and the free particle goes away from it. Finally, for velocities larger than this critical velocity (defined in Eq.~\ref{eq:vcr}), the free particle goes beyond the oscillator with the same velocity that it came with. In fact, this happens for any value of the mass ratio as shown in Figs.~\ref{Fig:slow_fast_osc}. 

With $m_{f} = m_{a}$, Eq.~(\ref{eq:velpcol2}) gives $v_{f}(t_{2}^+) = -v_{f}(0)$, that is, the free particle is reflected back with exactly the same energy as it came in. This is also corroborated by the numerical simulations in Fig.~\ref{Fig:slow_fast_osc}~(middle). The free particle and the oscillator interact twice in this region. At the first collision, the free particle transfers its entire energy to the oscillator, which is transferred back to it in the second collision. Such behavior persist even for relatively larger values of the oscillator frequency.

The two collision model presented here is not expected to work at higher frequencies, since the energy exchange between the two particles cannot be treated as well separated events. However, the approach still provides important insight into the energy exchange between the two particles.

\subsection{Energy loss calculation}
	In the previous section, it was shown that the exchange of energy between the two particles depends on both the spring constant $k$ and  the relative value of their masses. In order to gain some insight into the role of the mass ratio we present a perturbative approach. Setting the position of the oscillating particle as the perturbation parameter, the Gaussian potential in the Eqs.~(\ref{Eq:newton_p}) and (\ref{Eq:newton_osc}) can be expanded up to the first order in $x_{a}$. This gives
\ben
\begin{split}
\label{Eq:expansion}
(x_{f} - x_{a})e^{-\frac{(x_{f} - x_{a})^{2}}{\sigma^{2}}} &= \\
 x_{f}e^{-{x_{f}}^2/ \sigma^2} - & x_{a} (1- 
 \frac{2x_{f}^2}{\sigma^2}) e^{-x_{f}^2/\sigma^2} \;.
\end{split}
\een
Under this approximation, Eqs.~(\ref{Eq:newton_p}) and (\ref{Eq:newton_osc}) can be written as,
\ben
\begin{split}
\label{Eq:xfexpansion}
  \ddot{x}_{f} & = 2c_{f}x_{f}e^{-x_{f}^2/\sigma^2} &\\&
 - 2 c_{f} x_{a}(1-2x_{f}^2/\sigma^2)e^{-x_{f}^2/\sigma^2}\\
\label{Eq:xaexpansion} 
\ddot{x}_{a} &= -\frac{k}{m_{a}}x_{a} - 2 c_{a} x_{f}e^{-x_{f}^2/\sigma^2}\;,
\end{split}
\een  
where $c_{f} = \frac{U_{0}}{m_{f} \sigma^2}$ and $c_{a} = \frac{U_{0}}{m_{a} \sigma^2}$. The free particle moves under the influence of a static potential and the interaction appears as a perturbation linear in the oscillator coordinate ($x_{a}$). Computing for the work done by the free particle,
\ben \
\frac{dE}{dt} = F \dot {x}_{f} = m_{f} \ddot {x}_{f} \dot {x}_{f} \;,
\een
the following expression is obtained:
\ben
\begin{split}
\label{Eq:derivative}
\frac{dE}{dt} & = -2 c_{f} m_{f}  \dot {x}_{f} x_{a}(1 - 2x_{f}^2/\sigma^2)e^{-x_{f}^2/\sigma^2}  
\\ & + 2c_{f} m_{f} x_{f} \dot{x}_{f}e^{-x_{f}^2/\sigma^2}\;.
\end{split}
\een
The second term on the right hand side of Eq.~(\ref{Eq:derivative}) changes sign as $x_f$ goes from $+\infty$ to $-\infty$, and hence will be disregarded. The force due to the static potential is $F(x_{f}(t)) = c_f m_{f} x_{f}e^{-x_{f}^2/\sigma^2} $ whose derivative with respect to time is $\frac{dF}{dt} = c_f m_{f}\dot{x}_{f}(1 - 2x_{f}^2/\sigma^2)e^{-{x_{f}^2}/\sigma^2}$. The rate of loss of energy can be expressed in term of $\frac{dF}{dt}$ as, 
\ben \frac{dE}{dt} &= -2 x_{a} \frac{dF}{dt}
\een
and assuming that the free particle is completely transmitted with no rebound, the total energy loss is,

\ben
\begin{multlined}
 \Delta E = \int_0^{\infty} dt  \frac{dE}{dt} = -2
 \int_0^{t}dtx_{a}\frac{dF}{dt} = \\  2  \int_0^{\infty}
 dtF(x_{f}(t))\frac{dx_{a}}{dt}\;.
\end{multlined}
\label{dE}
\een
Assuming $ x_{a}F(t) \rightarrow 0 $ as $t\rightarrow \infty$, which is true for a pulse, and then using the formal solution for $x_a(t)$ from Eq.~(\ref{Eq:xaexpansion})

\ben \frac{dx_{a}}{dt} = -\frac{2}{m_a} \int_0^{t}
dsF(x_{f}(s))\cos(\omega(t-s))
\een
in Eq.~\ref{dE}, the energy lost by the free particle is

\begin{figure}[!h]
\centering
\includegraphics[width = 1.\columnwidth]{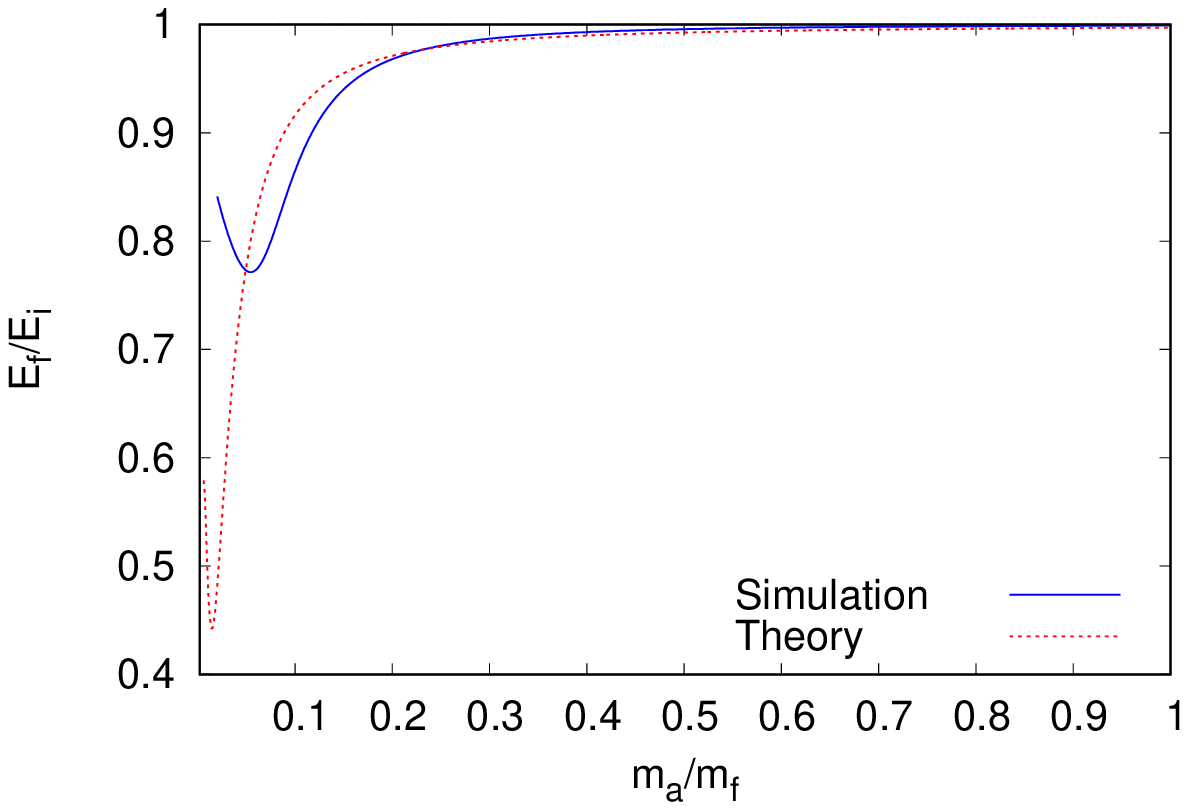}
\includegraphics[width = 1.\columnwidth]{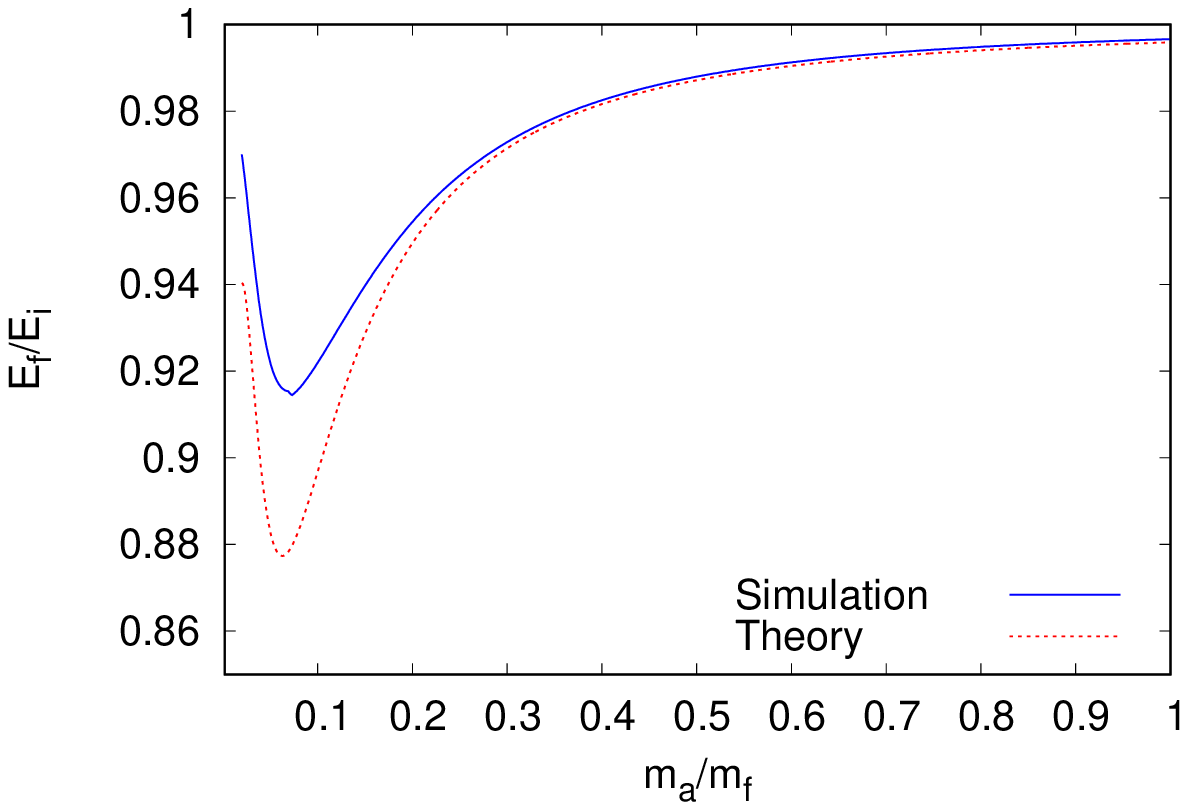}
\caption{Comparison of energy ratio ($E_f/E_i$) obtained from numerical simulation (solid line) and approximate solution from (Eq.~\ref{Eq:energyvariation}) (dashed line). $\sigma = 1$ in the top panel and equals $2$ in the bottom panel. The initial velocity of the free particle is $|v_f(0)|=4$ and all other parameters are equal to 1.}\label{fig:energyabs}
\end{figure}

\ben
\begin{split}
\label{Eq:energyvariation}
 \Delta E =&
 \\ -\frac{4}{m_a}&\int_0^{\infty}dtF(x_{f}(t))\int_0^{t}dsF(x_{f}(s))cos(\omega(t-s))\;.
\end{split}
\een
Fig.~\ref{fig:energyabs} shows the comparison between the exact solution (continuous line) and the energy ratio obtained by numerically solving Eq.~(\ref{Eq:energyvariation}). $F(x)$ is calculated by taking the solution of $x$ from Eq.~(\ref{Eq:xfexpansion}). The initial velocity of the free particle is kept sufficiently large ($|v_f(0)|=4$) to ensure that the free particle is always transmitted. As observed in Fig.~\ref{fig:energyabs}, for small values of $\sigma$, the perturbative approach is unable to provide a quantitative estimate of the velocity and location of the minima, however, for higher $\sigma$ values the agreement improves and the perturbative approach is able to predict accurately the location of the minima. Independent of the value of $\sigma$ the approach always provides a reasonable qualitative agreement. Fig.~\ref{fig:energyabs} shows that the energy ratio has a minimum at a relatively small value of the mass ratio ($m_{f} >> m_{a}$), implying that the energy loss increases as the mass of oscillator becomes smaller than the mass of the free particle up to a value beyond which it decreases. This behavior is evident since no energy loss is expected in the limit $m_{a}\rightarrow 0$.

\section{From one to $N$ particles}
The results presented in the previous sections correspond to the two particles model shown in Fig.~\ref{fig:model}. Remarkably, with such a simple first principle model we have shown that an effective dissipation emerges from the point of view of the incoming particle, which we postulate, as one of the possible mechanisms involved in the macroscopic friction.

\begin{figure}[!h]
	\centering
	\includegraphics[width = 1.\columnwidth]{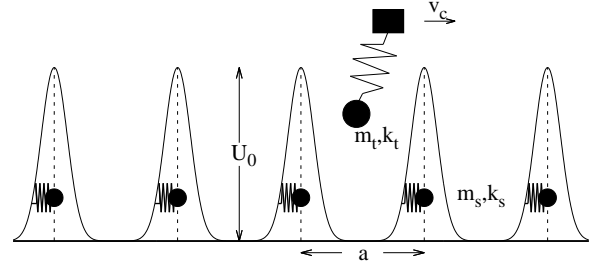}
	\caption{Extension of the two particles interaction model to a $N$-particle model: $m_{s}$ and $k_{s}$ are the mass and spring constant of the substrate; $m_{t}$, $k_{t}$ are the mass and spring constant of the tip. $U_{0}$ is the height of the interaction potential and $a$ is the lattice constant. The tip is driven by a support that moves at constant velocity $v_{c}$ .}\label{fig:model2}
\end{figure}

In order to test further such interpretations we include in this section some preliminary results in a more complex system: instead of a single bounded particle, a one-dimensional chain of independent oscillators, separated by a length $a=1$; and instead of a free particle, a tip mass connected by a spring of elastic constant $k_{t}$ to a support moving at constant velocity $v_{c}$ (see Fig.~\ref{fig:model2}).

The friction force $F_x$ is evaluated through the energy accumulated by the substrate, equating the rate of energy exchange  with the work done per unit time by the friction force in the steady regime:  
\ben
F_{x} v_{c} = \frac{dE}{dt} 
\label{Eq:FbyE}
\een
Since there is no ad hoc dissipation, the energy in the substrate increases linearly. Performing a linear regression of the substrate energy and using Eq.~(\ref{Eq:FbyE}), we obtain the frictional force $F_x$. 
Considering the work done by such force while the tip moves along the length $a=1$, we have that $F_x$ should be related with the 
energy exchange ($\Delta E$) in the two particles system studied in the previous sections, {\it i.e.}
\ben
F_x \approx \Delta E
\label{F_E}
\een

\begin{figure}[!h]
	\centering
	\includegraphics[width = 1.\columnwidth]{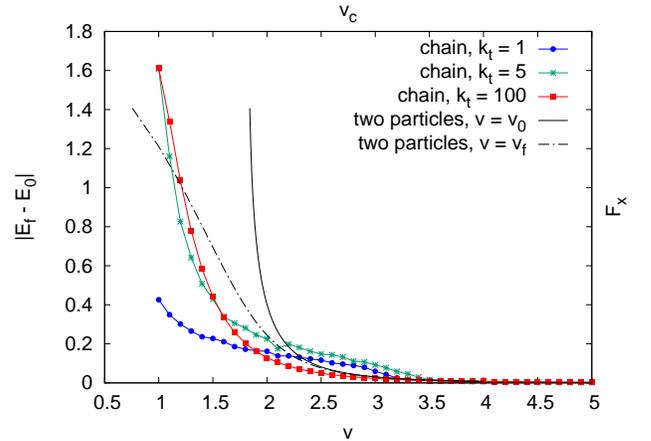}
	\caption{Absolute energy loss $|E_{f}-E_{0}|$ as a function of initial and final velocities for the two particle system (dashed and solid line) and friction force $F_{x}$ as a function of the cantilever velocity $v_{c}$ for the tip-substrate interaction. The effect of three different stiffness $k_{t}$ is presented, while the rest of the parameters are set to be equal to $1$.}\label{fig:one-N}
\end{figure}
Fig.~\ref{fig:one-N} shows the results of the numerical simulations
for the tip-substrate interaction (line and symbols) and for the two
particle interaction (dashed and solid line)~\footnote{This figure is part of a work in
	progress~\cite{Iglesias22017} where a extensive study over all the
	parameters of the model was conducted.}. For comparison, results
were obtained by changing the value of $k_{t}$, while the rest of the
parameters was kept equal to $1$.

The results show that for larger values of $k_{t}$, the particle behaves as a free particle sliding over a substrate without any constraint. For $k_{t}$ equals $100$, $F_{x}$ shows the same behavior as the energy loss observed in the case of the simplified two particle system. It can also be noticed that the curve is dislocated to the left, that is, the force is much greater in the chain for lower tip velocities. This is due to two reasons, the first is that the tip particle always moves at the same constant speed (unlike the first model, where it was thrown with an initial velocity) and the second is that throughout its course it interacts with all the particles of the substrate.

These results further consolidate the main idea of this contribution and demonstrate that the transfer of energy from the incoming particle to the confined one can be regarded as the emerging dissipation force of the system.

\section{Conclusion}
In an attempt to study the emergence of macroscopic friction from
microscopic dynamics of a simple system, we have presented results of
numerical experiments on a one-dimensional Hamiltonian system of two
particles: one with initial velocity thrown against other confined in
a harmonic potential. The interaction between them is repulsive and
short-ranged. This simple classical model exhibits features
characteristics of macroscopic effective friction.

Unlike in previous contributions to the study of the microscopic origin of friction, no ad hoc dissipative term is included in the model; the two particle system is conservative. This is the novelty in the present work. Despite its simplicity, the model shows effective dissipation on the free particle movement as a result of the interaction with the bound one. The free particle, that can be regarded as a model simplification of an adsorbate or a tip, moves past the bound particle (a simplification of a substrate) with less energy than it had before the interaction. The energy goes to the bound particle which remains oscillating after the interaction, representative of the heating of the substrate. The non-linearity of the potential plus the constraint in the bound particle are the key elements for this emergent behavior to appear. 

Apart from this general consequence, a range of rich and non-trivial results are obtained from the model. The energy exchange for three different mass ratios, as well as for different stiffness of the oscillator, was extensively analyzed. A theoretical reasoning to quantify the loss of energy of the free particle as a result of the interaction was proposed. For a certain range of incoming velocities, large losses of energy were observed. And as expected, no energy absorption occurs for large velocities of the incoming particle. 

With the present contribution, we hope to bring some physical comprehension of the microscopic mechanisms of energy dissipation between two particles which provide a simple model of the interaction between the substrate and the adsorbate or the tip of the microscope. Finally, we have also presented preliminary results with an extended version of the model in a periodic substrates showing emergence of friction in the steady regime for a sliding particle, instead of a ballistic one. This provides a step further in the understanding of the origin of macroscopic friction.

\section{Acknowledgments}
This work was supported by the Centro Latinoamericano de Física (CLAF), the Conselho Nacional de Desenvolvimento Científico e Tecnológico
(CNPq, Brasil), and in part by the Coordenação de Aperfeiçõamento de Pessoal de Nível Superior - Brasil (CAPES) - Finance Code 001.


\begin{thebibliography}{15}
\providecommand{\natexlab}[1]{#1}
\providecommand{\url}[1]{\texttt{#1}}
\expandafter\ifx\csname urlstyle\endcsname\relax
  \providecommand{\doi}[1]{doi: #1}\else
  \providecommand{\doi}{doi: \begingroup \urlstyle{rm}\Url}\fi

\bibitem[Krim(2002)]{Krim2002}
J. Krim.
\newblock Surface science and the atomic-scale origins of friction: what once
  was old is new again.
\newblock \emph{Surface Science}, 500\penalty0 (1–3):\penalty0 741 -- 758,
  2002.
\newblock ISSN 0039-6028.
\newblock \doi{https://doi.org/10.1016/S0039-6028(01)01529-1}.
\newblock URL
  \url{http://www.sciencedirect.com/science/article/pii/S0039602801015291}.

\bibitem[Bennewitz(2005)]{BENNEWITZ200542}
R. Bennewitz.
\newblock Friction force microscopy.
\newblock \emph{Materials Today}, 8\penalty0 (5):\penalty0 42 -- 48, 2005.
\newblock ISSN 1369-7021.
\newblock \doi{http://dx.doi.org/10.1016/S1369-7021(05)00845-X}.
\newblock URL
  \url{http://www.sciencedirect.com/science/article/pii/S136970210500845X}.

\bibitem[Buldum and Ciraci(1997)]{Buldum1997}
A.~Buldum and S.~Ciraci.
\newblock Atomic-scale study of dry sliding friction.
\newblock \emph{Phys. Rev. B}, 55(4):\penalty0 2606--2611, Jan 1997.
\newblock \doi{10.1103/PhysRevB.55.2606}.
\newblock URL \url{http://link.aps.org/doi/10.1103/PhysRevB.55.2606}.

\bibitem[Persson(2000)]{Persson2000}
B.~N.~J. Persson.
\newblock \emph{{Sliding Friction}}.
\newblock Springer Berlin Heidelberg, 2000.
\newblock \doi{10.1007/978-3-662-04283-0}.
\newblock URL \url{http://dx.doi.org/10.1007/978-3-662-04283-0}.

\bibitem[Holscher et~al.(2008)Holscher, Schirmeisen, and Schwarz]{Holscher2008}
H.~Holscher, A.~Schirmeisen, and U.~D Schwarz.
\newblock {Principles of atomic friction: from sticking atoms to superlubric
  sliding}.
\newblock \emph{Philosophical Transactions of the Royal Society A: Mathematical
  Physical and Engineering Sciences}, 366\penalty0 (1869):\penalty0 1383--1404,
  apr 2008.
\newblock \doi{10.1098/rsta.2007.2164}.
\newblock URL \url{http://dx.doi.org/10.1098/rsta.2007.2164}.

\bibitem[Tomlinson(1929)]{Tomlinson1929}
G. A. Tomlinson.
\newblock {{CVI}. A molecular theory of friction}.
\newblock \emph{The London Edinburgh, and Dublin Philosophical Magazine and
  Journal of Science}, 7\penalty0 (46):\penalty0 905--939, jun 1929.
\newblock \doi{10.1080/14786440608564819}.
\newblock URL \url{http://dx.doi.org/10.1080/14786440608564819}.

\bibitem{Prandtl1928}
L. Prandtl, 
Ein Gedankenmodell zur kinetischen Theorie der festen Körper.
ZAMM 8, 85–106,1928.

\bibitem{Prandtl1904}
L. Prandtl,
Über Flüssigkeitsbewegung bei sehr kleiner Reibung, Verhandlungen III,
p. 484,
Intern. Math. Kongress, Heidelberg,1904.

\bibitem{Prandtl1920}
L. Prandtl,
Über die Härte plastischer Körper,
Nachrichten Göttinger Akad. Wiss,1920.

\bibitem[Kontorova and Frenkel(1938)]{Kontorova:431596}
T.~Kontorova and J.~Frenkel.
\newblock {On the theory of plastic deformation and twinning. II.}
\newblock \emph{Zh. Eksp. Teor. Fiz.}, 8:\penalty0 1340--1348, 1938.
\newblock URL \url{http://cds.cern.ch/record/431596}.

\bibitem[Frenkel and Kontorova(1939)]{Frenkel:431595}
J.~Frenkel and T.~Kontorova.
\newblock {On the theory of plastic deformation and twinning}.
\newblock \emph{Izv. Akad. Nauk, Ser. Fiz.}, 1:\penalty0 137--149, 1939.
\newblock URL \url{http://cds.cern.ch/record/431595}.

\bibitem[Braun and Kivshar(2004)]{braun2004frenkel}
O. M. Braun and Y. S. Kivshar.
\newblock \emph{The Frenkel-Kontorova Model: Concepts, Methods, and
  Applications}.
\newblock Physics and Astronomy Online Library. Berlin :Springer, 2004.
\newblock ISBN 9783540407713.
\newblock URL \url{https://books.google.com.br/books?id=zyoT068mu0YC}.

\bibitem[Gon\ifmmode~\mbox{\c{c}}\else \c{c}\fi{}alves
  et~al.(2004)Gon\ifmmode~\mbox{\c{c}}\else \c{c}\fi{}alves, Kenkre, and
  Bishop]{Goncalves2004}
S.~Gon\ifmmode~\mbox{\c{c}}\else \c{c}\fi{}alves, V.~M. Kenkre, and A.~R.
  Bishop.
\newblock Nonlinear friction of a damped dimer sliding on a periodic substrate.
\newblock \emph{Phys. Rev. B}, 70(19):\penalty0 195415, Nov 2004.
\newblock \doi{10.1103/PhysRevB.70.195415}.
\newblock URL \url{https://link.aps.org/doi/10.1103/PhysRevB.70.195415}.

\bibitem[Gon\ifmmode~\mbox{\c{c}}\else \c{c}\fi{}alves
  et~al.(2005)Gon\ifmmode~\mbox{\c{c}}\else \c{c}\fi{}alves, Fusco, Bishop, and
  Kenkre]{Goncalves2005}
S.~Gon\ifmmode~\mbox{\c{c}}\else \c{c}\fi{}alves, C.~Fusco, A.~R. Bishop, and
  V.~M. Kenkre.
\newblock Bistability and hysteresis in the sliding friction of a dimer.
\newblock \emph{Phys. Rev. B}, 72(19):\penalty0 195418, Nov 2005.
\newblock \doi{10.1103/PhysRevB.72.195418}.
\newblock URL \url{https://link.aps.org/doi/10.1103/PhysRevB.72.195418}.

\bibitem[Fusco and Fasolino(2005)]{Fusco2005}
C.~Fusco and A.~Fasolino.
\newblock Velocity dependence of atomic-scale friction: A comparative study of
  the one- and two-dimensional Tomlinson model.
\newblock \emph{Phys. Rev. B}, 71(4):\penalty0 045413, Jan 2005.
\newblock \doi{10.1103/PhysRevB.71.045413}.
\newblock URL \url{https://link.aps.org/doi/10.1103/PhysRevB.71.045413}.

\bibitem[Tiwari et~al.(2008)Tiwari, Gon{\c{c}}alves, and Kenkre]{Tiwari2008}
M.~Tiwari, S.~Gon{\c{c}}alves, and V.~M. Kenkre.
\newblock {Generalization of a nonlinear friction relation for a dimer sliding
  on a periodic substrate}.
\newblock \emph{The European Physical Journal B}, 62\penalty0 (4):\penalty0
  459--464, apr 2008.
\newblock \doi{10.1140/epjb/e2008-00194-9}.
\newblock URL \url{http://dx.doi.org/10.1140/epjb/e2008-00194-9}.

\bibitem[Neide et~al.(2010)Neide, Kenkre, and Gon\ifmmode~\mbox{\c{c}}\else
  \c{c}\fi{}alves]{Neide2010}
I.~G. Neide, V.~M. Kenkre, and S.~Gon\ifmmode~\mbox{\c{c}}\else
  \c{c}\fi{}alves.
\newblock Effects of rotation on the nonlinear friction of a damped dimer
  sliding on a periodic substrate.
\newblock \emph{Phys. Rev. E}, 82 \penalty0 (4): \penalty0 046601, Oct 2010.
\newblock \doi{10.1103/PhysRevE.82.046601}.
\newblock URL \url{https://link.aps.org/doi/10.1103/PhysRevE.82.046601}.


\bibitem[Apostoli(2017)]{Apostoli2017}
C. Apostoli, G. Giusti, J. Ciccoianni, G. Riva, R. Capozza, R. L. Woulaché, A. Vanossi, E. Panizon, N. Manini.
Velocity dependence of sliding friction on a crystalline surface.  
\newblock \emph{Beilstein J.Nanotechnol.}, 8:\penalty0 2186–2199, 2017.
\newblock \doi{10.3762/bjnano.8.218}.

 Beilstein J. Nanotechnol. 2017, 8, 2186–2199. doi:10.3762/bjnano.8.218


\bibitem[Secrest(1969)]{Secrest1969}
D. Secrest.
\newblock Linear collision of a classical harmonic oscillator with a particle
  at high energies.
\newblock \emph{The Journal of Chemical Physics}, 51\penalty0 (1):\penalty0
  421--425, 1969.
\newblock \doi{10.1063/1.1671741}.
\newblock URL \url{https://doi.org/10.1063/1.1671741}.

\bibitem[Storm and Thiele(1973)]{Storm1973}
D. Storm and E. Thiele.
\newblock Collinear collision between a particle and a harmonic oscillator with
  a Morse potential interaction.
\newblock \emph{The Journal of Chemical Physics}, 59\penalty0 (6):\penalty0
  3313--3318, 1973.
\newblock \doi{10.1063/1.1680475}.
\newblock URL \url{https://doi.org/10.1063/1.1680475}.

\bibitem[Teubner(2005)]{Teubner2005}
M. Teubner.
\newblock Linear collision of a classical harmonic oscillator with a mass point
  in the high-frequency region.
\newblock \emph{Phys. Rev. A}, 72(4):\penalty0 042703, Oct 2005.
\newblock \doi{10.1103/PhysRevA.72.042703}.
\newblock URL \url{https://link.aps.org/doi/10.1103/PhysRevA.72.042703}.

\bibitem[Mavri et~al.(1994)Mavri, Lensink, and Berendsen]{Mavri1994}
J. Mavri, M. Lensink, and H.~J.C. Berendsen.
\newblock Treatment of inelastic collisions of a particle with a quantum
  harmonic oscillator by density matrix evolution.
\newblock \emph{Molecular Physics}, 82\penalty0 (6):\penalty0 1249--1257, 1994.
\newblock \doi{10.1080/00268979400100884}.
\newblock URL \url{https://doi.org/10.1080/00268979400100884}.

\bibitem[Iglesias(2017)]{Iglesias22017}
M. L. Iglesias and S. Gonçalves.
\newblock Tomlinson model improved with no ad hoc dissipation,2018
\url{arXiv:1708.03415} work in progress


\end{thebibliography}
\end{document}